\documentclass[twoside,twocolumn,9pt]{article}
\usepackage{extsizes}
\usepackage[super,sort&compress,comma,numbers]{natbib} 
\usepackage[version=3]{mhchem}
\usepackage[left=1.5cm, right=1.5cm, top=1.785cm, bottom=2.0cm]{geometry}
\usepackage{balance}
\usepackage{times,mathptmx}
\usepackage{sectsty}
\usepackage{graphicx} 
\usepackage{lastpage}
\usepackage[format=plain,justification=justified,singlelinecheck=false,font={stretch=1.125,small,sf},labelfont=bf,labelsep=space]{caption}
\usepackage{float}
\usepackage{fancyhdr}
\usepackage{fnpos}
\usepackage{array}
\usepackage{droidsans}
\usepackage{charter}
\usepackage[T1]{fontenc}
\usepackage[usenames,dvipsnames]{xcolor}
\usepackage{setspace}
\usepackage[compact]{titlesec}
\usepackage{hyperref}
\usepackage{xspace}

\usepackage{epstopdf}

\DeclareMathAlphabet{\mathpzc}{OT1}{pzc}{m}{it}
\DeclareSymbolFont{matha}{OML}{txmi}{m}{it}
\DeclareMathSymbol{\varv}{\mathord}{matha}{118}
\usepackage{dcolumn}
\usepackage{xspace}
\newcolumntype{.}{D{.}{.}{-1}}
\newcolumntype{d}[1]{D{.}{.}{#1}}

\definecolor{cream}{RGB}{222,217,201}

\begin{document}

\pagestyle{fancy}
\thispagestyle{plain}
\fancypagestyle{plain}{

\renewcommand{\headrulewidth}{0pt}
}

\makeFNbottom
\makeatletter
\renewcommand\LARGE{\@setfontsize\LARGE{15pt}{17}}
\renewcommand\Large{\@setfontsize\Large{12pt}{14}}
\renewcommand\large{\@setfontsize\large{10pt}{12}}
\renewcommand\footnotesize{\@setfontsize\footnotesize{7pt}{10}}
\makeatother

\renewcommand{\thefootnote}{\fnsymbol{footnote}}
\renewcommand\footnoterule{\vspace*{1pt}%
\color{cream}\hrule width 3.5in height 0.4pt \color{black}\vspace*{5pt}} 
\setcounter{secnumdepth}{5}

\makeatletter 
\renewcommand\@biblabel[1]{#1}
\renewcommand\@makefntext[1]%
{\noindent\makebox[0pt][r]{\@thefnmark\,}#1}
\makeatother 
\renewcommand{\figurename}{\small{Fig.}~}
\sectionfont{\sffamily\Large}
\subsectionfont{\normalsize}
\subsubsectionfont{\bf}
\setstretch{1.125} 
\setlength{\skip\footins}{0.8cm}
\setlength{\footnotesep}{0.25cm}
\setlength{\jot}{10pt}
\titlespacing*{\section}{0pt}{4pt}{4pt}
\titlespacing*{\subsection}{0pt}{15pt}{1pt}

\fancyfoot{}
\fancyfoot[RO]{\footnotesize{\sffamily{1--\pageref{LastPage} ~\textbar  \hspace{2pt}\thepage}}}
\fancyfoot[LE]{\footnotesize{\sffamily{\thepage~\textbar\hspace{3.45cm} 1--\pageref{LastPage}}}}
\fancyhead{}
\renewcommand{\headrulewidth}{0pt} 
\renewcommand{\footrulewidth}{0pt}
\setlength{\arrayrulewidth}{1pt}
\setlength{\columnsep}{6.5mm}
\setlength\bibsep{1pt}

\makeatletter 
\newlength{\figrulesep} 
\setlength{\figrulesep}{0.5\textfloatsep} 

\newcommand{\topfigrule}{\vspace*{-1pt}%
\noindent{\color{cream}\rule[-\figrulesep]{\columnwidth}{1.5pt}} }

\newcommand{\botfigrule}{\vspace*{-2pt}%
\noindent{\color{cream}\rule[\figrulesep]{\columnwidth}{1.5pt}} }

\newcommand{\dblfigrule}{\vspace*{-1pt}%
\noindent{\color{cream}\rule[-\figrulesep]{\textwidth}{1.5pt}} }

\makeatother

\providecommand*{\iu}{\ensuremath{\mathrm{i}}}
\providecommand*{\diff}{\ensuremath{\mathrm{d}}}
\providecommand*{\ee}{\ensuremath{\mathrm{e}}}

\providecommand*{\wn}{cm$^{-1}$\xspace}
\providecommand*{\hms}[3]{\ensuremath{{#1}^\mrm{h}{#2}^\mrm{m}{#3}^\mrm{s}}\xspace}
\providecommand*{\dms}[3]{\ensuremath{#1^\circ #2^\prime #3^{\prime\prime}}\xspace}

\newcommand{\qnd}{\ensuremath{^{15}}ND\xspace}
\newcommand{\qnh}{\ensuremath{^{15}}NH\xspace}
\newcommand{\elgs}{\ensuremath{X\,^3\Sigma^-}\xspace}
\newcommand{\elex}{\ensuremath{A\,^3\Pi}\xspace}
\newcommand{\mrm}[1]{\ensuremath{\mathrm{#1}}}
\newcommand{\mcl}[3]{\multicolumn{#1}{#2}{#3}}
\newcommand{\mrw}[3]{\multirow{#1}{#2}{#3}}
\newcommand{\pha}[1]{\phantom{#1}}
\newcommand{\vet}[1]{\ensuremath{\boldsymbol{#1}}}
\newcommand{\varY}{\mathpzc{Y}}

\twocolumn[
  \begin{@twocolumnfalse}
\vspace{3cm}
\sffamily
\begin{tabular}{m{4.5cm} p{13.5cm}}

\includegraphics{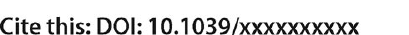}   &
\noindent\LARGE{\textbf{%
  The rotational spectrum of $^{15}$ND. Isotopic-independent Dunham-type 
  analysis of the imidogen radical$^\dag$
  }}                              \\
\vspace{0.3cm} & \vspace{0.3cm}   \\
                                  &
\noindent\large{Mattia Melosso\textit{$^{a}$}, 
                Luca Bizzocchi\textit{$^{b,\ast}$}, 
                Filippo Tamassia\textit{$^{c}$},
                Claudio Degli Esposti\textit{$^{a}$},
                Elisabetta Can\`{e}\textit{$^{c}$}, 
                and Luca Dore\textit{$^{a,\ast}$}
                }                 \\

\includegraphics{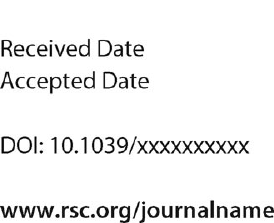} & 
\noindent\normalsize{%
The rotational spectrum of \qnd in its ground electronic \elgs state has been observed 
for the first time. 
Forty-three hyperfine-structure components belonging to the ground and $\varv = 1$ vibrational
states have been recorded with a frequency-modulation millimeter-/submillimeter-wave 
spectrometer. 
These new measurements, together with the ones available for the other isotopologues NH, ND, 
and \qnh, have been simultaneously analysed using the Dunham model to represent the 
ro-vibrational, fine, and hyperfine energy contributions.
The least-squares fit of more than 1500 transitions yielded an extensive set of isotopically 
independent $U_{lm}$ parameters plus 13 Born--Oppenheimer Breakdown coefficients $\Delta_{lm}$.
As an alternative approach, we performed a Dunham analysis in terms of the most abundant 
isotopologue coefficients $Y_{lm}$ and some isotopically dependent Born--Oppenheimer Breakdown 
constants $\delta_{lm}$ [R.~J. Le Roy, \textit{J. Mol. Spectrosc.} \textbf{194}, 189 (1999)].
The two fits provide results of equivalent quality. 
The Born--Oppenheimer equilibrium bond distance for the imidogen radical has been calculated 
[$r_e^\mrm{BO} =103.606721(13)$ pm] and zero point energies have been derived for all the 
isotopologues.
} \\

\end{tabular}
\end{@twocolumnfalse} \vspace{0.6cm}
]

\renewcommand*\rmdefault{bch}\normalfont\upshape
\rmfamily
\section*{}
\vspace{-1cm}

\footnotetext{$^\ast$~Corresponding author.\\
\textit{$^{a}$~Dipartimento di Chimica ``Giacomo Ciamician'', 
                             Universit\`a di Bologna,
                             Via F.~Selmi 2, 40126 Bologna (Italy).
                             E-mail: mattia.melosso2@unibo.it,
                                     claudio.degliesposti@unibo.it
                                     luca.dore@unibo.it}}
\footnotetext{\textit{$^{b}$~Center for Astrochemical Studies, 
                             Max-Planck-Institut f\"ur extraterrestrische Physik, 
                             Gie\ss enbachstr.~1, 85748 Garching bei M\"unchen (Germany)
                             E-mail: bizzocchi@mpe.mpg.de}}
\footnotetext{\textit{$^{c}$~Dipartimento di Chimica Industriale ``Toso Montanari'', 
                             Universit\`a di Bologna,
                             Viale del Risorgimento~4, 40136 Bologna (Italy).
                             E-mail: filippo.tamassia@unibo.it,
                         			 elisabetta.cane@unibo.it}}


\footnotetext{\dag~Electronic Supplementary Information (ESI) available: 
              The .LIN and .PAR files for the SPFIT programm are provided for both 
              the single-species and multi-isotopologues fits.
              A reformatted list of all the transitions used in the Dunham-type analysis,
              together with their residuals from the final fit, is also included as ESI.
              See DOI: 10.1039/c8cp04498h}
              



\section{Introduction}
\indent\indent
The imidogen radical has been the subject of many spectroscopic, computational and 
astrophysical studies.
This diatomic radical belongs to the first-row hydrides, is commonly observed in the 
combustion products of nitrogen-bearing compounds \cite{alexander1991,miller1989}, 
and is also an intermediate in the formation process of ammonia in the interstellar 
medium (ISM) \cite{Wagenblast1993}.
The main isotopologue of imidogen, NH, has been detected in a wide-variety of 
environments, from the Earths' atmosphere to astronomical objects, such as comets 
\cite{Feldman1993}, many types of stars \cite{Schimtt1969,Ridgway1984} including the Sun 
\cite{Grevesse1990,Geller1991}, diffuse clouds \cite{Meyer1991}, massive star-forming (SF)
regions \cite{Persson2010} and, very recently, in prestellar cores \cite{Bacmann2016}.
Also its deuterated counterpart ND has been identified in the ISM, towards the 
young solar-mass protostar IRAS16293 \cite{Bacmann2010} and in the prestellar core 
16293E \cite{Bacmann2016}. \\
A lot of studies have been devoted to the origin of interstellar imidogen and different 
formation models have been proposed to explain its observed abundance in various 
sources.
Two main formation routes have been devised for the NH radical: from the electronic 
recombination of NH$^+$ and NH$_{2}^+$, intermediates in the synthesis of 
interstellar ammonia \cite{Herbst1987,Galloway1989} starting with N$^+$ or, alternatively, 
via dissociative recombination of N$_{2}$H$^+$ \cite{Geppert2004}.
However, the mechanism of NH production in the ISM is still debated \cite{Persson2010},
and grain-surface processes might also play a significant role \cite{Oneill2002}.
Imidogen, together with other light hydrides, often appears in the first steps of chemical 
networks leading to more complex N-bearing molecules. Its observation thus provides 
crucial constraints for the chemical modeling of astrophysical sources \cite{LeGal2014}.
Also the rare isotopologues of this radical yield important astrochemical insights.
Being proxies for N and D isotopic fractionation processes, they may help to trace the 
evolution of gas and dust during the star formation, thus shedding light on the link between 
Solar System materials and the parent ISM \cite{Aleon2010}.
This is particularly relevant for nitrogen, whose molecular isotopic compositions exhibits
large and still unexplained variations \cite{Marty2011,Furi2015}.
Measuring the isotopic ratios in imidogen provides useful complementary information 
on the already measured H/D, $^{14}$N/$^{15}$N in ammonia (including the $^{15}$NH$_2$D 
species \cite{Gerin2009}).

As far as the laboratory work is concerned, there is a substantial amount of spectroscopic
data for the most abundant species and less extensive measurements for \qnh and ND. 
A detailed description of the spectroscopic studies of imidogen can be found in the latest experimental works on NH \cite{Ram2010}, 
\qnh \cite{Bizzocchi2018}, and ND \cite{Dore2011}.
It has to be noticed that no experimental data or theoretical computations were available 
in literature for the doubly substituted species \qnd up to date.
In this work, we report the first observation of its pure rotational spectrum in the ground 
electronic state $X^3\Sigma^-$ recorded up to 1.068\,THz.
A limited number of new transition frequencies for the isotopologues NH and ND in the $v = 1$
excited state have also been measured in the course of the present investigation.
This new set of data, together with the literature data for NH, \qnh
and ND, have been analysed in a global multi-isotopologue fit to give a comprehensive
set of isotopically independent spectroscopic parameters.
Thanks to the high precision of the measurements, several Born--Oppenheimer Breakdown (BOB) 
constants ($\Delta_{lm}$) could be determined from a Dunham-type analysis.
The alternative Dunham approach proposed by \citet{LeRoy1999} has been also employed. 
In this case, the results are expressed in terms of the parent species coefficients 
$Y_{lm}$ plus some isotopically dependent BOB constants ($\delta_{lm}$).

Finally, very accurate equilibrium bond distances $r_e$ (including the Born--Oppenheimer 
bond distance $r_e^{\mrm{BO}}$) and Zero-Point Energies (ZPE) for each isotopologue have
been computed from the determined spectroscopic constants.

\section{Experiments} \label{sec:exp}
\indent\indent
The rotational spectrum of \qnd radical in its ground vibronic state $X^3\Sigma^-$ 
has been recorded with a frequency-modulation millimeter-/submillimeter-wave spectrometer.
The primary source of radiation was constituted by a series of Gunn diodes 
(Radiometer Physics GmbH, J.~E. Carlstrom Co.) emitting in the range 80--134\,GHz, 
whose frequency is stabilized by a Phase-Lock-Loop (PLL) system. 
The PLL allowed the stabilization of the Gunn oscillator with respect to a frequency 
synthesizer (Schomandl ND~1000), which was driven by a 5\,MHz rubidium frequency standard. 
Higher frequencies were obtained by using passive multipliers (RPG, $\times 6$ and 
$\times 9$).
The frequency modulation of the output radiation was realized by sine-wave modulating at 
6\,kHz the reference signal of the wide-band Gunn synchronizer. 
The signal was detected by a liquid-helium-cooled InSb hot electron bolometer 
(QMC Instr.~Ltd. type QFI/2) and then demodulated at 2$f$ by a lock-in amplifier. 
The experimental uncertainties of present measurements are between 40 and 80\,kHz in most 
cases, up to 500 kHz for a few disturbed lines. \\
The $^{15}$ND radical was formed in a glow-discharge plasma with the same apparatus 
employed to produce other unstable and rare species
(e.g., ND$_2$\cite{Melosso2017} and $^{15}$N$_2$H$^+$\cite{dore2017doubly}).
The optimum production was attained in a DC discharge of a mixture of $^{15}$N$_2$ 
(5--7\,mTorr) and D$_2$ (1--2\,mTorr) in Ar as buffer gas (15\,mTorr). 
Typically, a voltage of 1\,kV and a current of 60\,mA were employed.
The absorption cell was cooled down at \textit{ca.}\ $-190^\circ$C by liquid-nitrogen 
circulation.
\bigskip

\begin{table}[t]
  \caption[]{Spectroscopic constants determined for \qnd in the ground and $\varv = 1$
             vibrational states.}
  \label{tab:15ndres}
  \centering\small
  \vspace{2ex}
  \begin{tabular}{ll d{8} d{8}}
   \hline\hline \\[-1ex]
     Constant  & \mcl{1}{c}{Unit} & \mcl{1}{c}{$v = 0$}  &  \mcl{1}{c}{$v = 1$} \\[0.5ex]
   \hline \\[-1ex]
     $B_v$          &  / MHz  &  261083.4809(96)  &  253597.797(24)   \\[0.5ex]
     $D_v$          &  / MHz  &     14.3906(13)   &     14.3906^a     \\[0.5ex]
     $\lambda_v$    &  / MHz  &   27544.852(22)   &  27544.852^a      \\[0.5ex]
     $\gamma_v$     &  / MHz  &    -876.139(15)   &    -841.674(46)   \\[0.5ex]
     $\gamma_{Nv}$  &  / MHz  &      0.1241(20)   &       0.1241^a    \\[0.5ex]
     $b_{F,v}(\mrm{^{15}N})$   &  / MHz  &     -26.519(20)   &     -25.944(41)   \\[0.5ex]
     $c_v(\mrm{^{15}N})$       &  / MHz  &       95.154(56)  &      94.48(30)    \\[0.5ex]
     $C_{I,v}(\mrm{^{15}N})$   &  / MHz  &      -0.124(14)   &      -0.124^a     \\[0.5ex]
     $b_{F,v}(\mrm{D})$   &  / MHz  &     -10.062(21)   &     -10.524(42)   \\[0.5ex]
     $c_v(\mrm{D})$       &  / MHz  &       14.236(78)  &      13.18(22)    \\[0.5ex]
     $eQq_v(\mrm{D})$     &  / MHz  &        0.271(93)   &       0.271^a      \\[0.5ex]
     \hline \\[-1ex]
     $\sigma_\mrm{w}^b$  &    &         \mcl{2}{c}{0.84}              \\[0.5ex]
     $rms$          &  / MHz  &         \mcl{2}{c}{0.080}             \\[0.5ex]
     no. of lines   &         &    \mcl{1}{c}{34}  & \mcl{1}{c}{9}    \\[0.5ex]
   \hline\hline \\
  \end{tabular}
  \begin{minipage}{0.45\textwidth}
    \textbf{Notes.} \\
    Number in parentheses are the $1\sigma$ statistical errors in unit of the last quoted digit. 
    $^{(a)}$ Parameter held fixed in the fit. $^{(b)}$ Fit standard deviation.
  \end{minipage}
\end{table}

\begin{table}[t]
	\caption[]{Observed frequencies and residuals (in MHz) from the single-isotopologue fit of \qnd in the ground and first
		vibrational excited states.}
	\label{tab:15ndfit}
	\centering\small
	\vspace{2ex}
	\scalebox{0.7}{
		\begin{tabular}{l cccc cccc d{8} d{8} d{8}}
			\hline\hline \\[-1ex]
			State & $N'$ & $J'$ & $F_1'$ & $F'$ & $N"$ & $J"$ & $F_1"$ & $F"$  & \mcl{1}{c}{Obs. Freq.} & \mcl{1}{c}{Obs.- Calc.}  &  \mcl{1}{c}{Rel. weight} \\[0.5ex]
			\hline \\[-1ex]
			$v=0$ & 1 & 0 & 0.5 & 1.5 & 0 & 1 & 0.5 & 0.5   &   487528.290(80)  &   0.119  & 0.95  \\
			& 1 & 0 & 0.5 & 0.5 & 0 & 1 & 0.5 & 0.5   &   487528.290(80)  &   0.119  & 0.05  \\
			& 1 & 0 & 0.5 & 1.5 & 0 & 1 & 0.5 & 1.5   &   487547.166(80)  &   0.012  & 0.79  \\
			& 1 & 0 & 0.5 & 0.5 & 0 & 1 & 0.5 & 1.5   &   487547.166(80)  &   0.012  & 0.21  \\
			& 1 & 0 & 0.5 & 1.5 & 0 & 1 & 1.5 & 1.5   &   487578.185(500) &  -0.229  & 0.21  \\
			& 1 & 0 & 0.5 & 0.5 & 0 & 1 & 1.5 & 1.5   &   487578.185(500) &  -0.229  & 0.79  \\
			& 1 & 0 & 0.5 & 1.5 & 0 & 1 & 1.5 & 2.5   &   487593.045(500) &  -0.332  &       \\
			& 1 & 2 & 2.5 & 3.5 & 0 & 1 & 1.5 & 2.5   &   517707.856(50)  &   0.002  &       \\
			& 1 & 2 & 2.5 & 2.5 & 0 & 1 & 1.5 & 1.5   &   517709.082(50)  &  -0.060  &       \\
			& 1 & 2 & 2.5 & 1.5 & 0 & 1 & 1.5 & 0.5   &   517709.837(50)  &   0.024  &       \\
			& 1 & 2 & 1.5 & 2.5 & 0 & 1 & 0.5 & 1.5   &   517712.966(50)  &  -0.008  &       \\
			& 1 & 2 & 2.5 & 1.5 & 0 & 1 & 1.5 & 1.5   &   517721.385(50)  &   0.046  &       \\
			& 1 & 2 & 2.5 & 2.5 & 0 & 1 & 1.5 & 2.5   &   517724.182(50)  &   0.052  &       \\
			& 1 & 2 & 1.5 & 1.5 & 0 & 1 & 0.5 & 1.5   &   517730.725(50)  &  -0.032  &       \\
			& 1 & 2 & 1.5 & 2.5 & 0 & 1 & 1.5 & 2.5   &   517759.241(50)  &   0.038  &       \\
			& 1 & 1 & 0.5 & 1.5 & 0 & 1 & 0.5 & 0.5   &   541723.559(80)  &  -0.009  & 0.85  \\
			& 1 & 1 & 0.5 & 0.5 & 0 & 1 & 0.5 & 0.5   &   541723.559(80)  &  -0.009  & 0.15  \\
			& 1 & 1 & 0.5 & 1.5 & 0 & 1 & 0.5 & 1.5   &   541742.758(80)  &  -0.015  & 0.43  \\
			& 1 & 1 & 0.5 & 0.5 & 0 & 1 & 0.5 & 1.5   &   541742.758(80)  &  -0.015  & 0.57  \\
			& 1 & 1 & 1.5 & 1.5 & 0 & 1 & 0.5 & 0.5   &   541751.550(80)  &   0.064  & 0.74  \\
			& 1 & 1 & 1.5 & 0.5 & 0 & 1 & 0.5 & 0.5   &   541751.550(80)  &   0.064  & 0.26  \\
			& 1 & 1 & 0.5 & 1.5 & 0 & 1 & 1.5 & 0.5   &   541762.512(80)  &  -0.042  & 0.31  \\
			& 1 & 1 & 0.5 & 0.5 & 0 & 1 & 1.5 & 0.5   &   541762.512(80)  &  -0.042  & 0.69  \\
			& 1 & 1 & 1.5 & 2.5 & 0 & 1 & 0.5 & 1.5   &   541769.817(80)  &  -0.122  & 0.98  \\
			& 1 & 1 & 1.5 & 1.5 & 0 & 1 & 0.5 & 1.5   &   541769.817(80)  &  -0.122  & 0.02  \\
			& 1 & 1 & 0.5 & 1.5 & 0 & 1 & 1.5 & 1.5   &   541773.823(80)  &   0.053  & 0.88  \\
			& 1 & 1 & 0.5 & 0.5 & 0 & 1 & 1.5 & 1.5   &   541773.823(80)  &   0.053  & 0.12  \\
			& 1 & 1 & 0.5 & 1.5 & 0 & 1 & 1.5 & 2.5   &   541788.538(80)  &  -0.154  &       \\
			& 1 & 1 & 1.5 & 1.5 & 0 & 1 & 1.5 & 0.5   &   541790.239(80)  &  -0.069  & 0.37  \\
			& 1 & 1 & 1.5 & 0.5 & 0 & 1 & 1.5 & 0.5   &   541790.239(80)  &  -0.069  & 0.63  \\
			& 1 & 1 & 1.5 & 2.5 & 0 & 1 & 1.5 & 1.5   &   541801.642(80)  &  -0.006  & 0.12  \\
			& 1 & 1 & 1.5 & 1.5 & 0 & 1 & 1.5 & 1.5   &   541801.642(80)  &  -0.006  & 0.64  \\
			& 1 & 1 & 1.5 & 0.5 & 0 & 1 & 1.5 & 1.5   &   541801.642(80)  &  -0.006  & 0.24  \\
			& 1 & 1 & 1.5 & 2.5 & 0 & 1 & 1.5 & 2.5   &   541816.257(80)  &   0.027  & 0.84  \\
			& 1 & 1 & 1.5 & 1.5 & 0 & 1 & 1.5 & 2.5   &   541816.257(80)  &   0.027  & 0.16  \\
			& 2 & 1 & 1.5 & 0.5 & 1 & 1 & 0.5 & 0.5   &  1009356.833(40)  &  -0.043  & 0.78  \\
			& 2 & 1 & 1.5 & 0.5 & 1 & 1 & 0.5 & 1.5   &  1009356.833(40)  &  -0.043  & 0.22  \\
			& 2 & 3 & 3.5 & 3.5 & 1 & 2 & 2.5 & 2.5   &  1041497.348(40)  &   0.028  & 0.33  \\
			& 2 & 3 & 3.5 & 4.5 & 1 & 2 & 2.5 & 3.5   &  1041497.348(40)  &   0.028  & 0.44  \\
			& 2 & 3 & 3.5 & 2.5 & 1 & 2 & 2.5 & 1.5   &  1041497.348(40)  &   0.028  & 0.24  \\
			& 2 & 3 & 2.5 & 2.5 & 1 & 2 & 1.5 & 1.5   &  1041499.205(40)  &  -0.006  & 0.31  \\
			& 2 & 3 & 2.5 & 3.5 & 1 & 2 & 1.5 & 2.5   &  1041499.205(40)  &  -0.006  & 0.50  \\
			& 2 & 3 & 2.5 & 1.5 & 1 & 2 & 1.5 & 0.5   &  1041499.205(40)  &  -0.006  & 0.19  \\
			& 2 & 3 & 3.5 & 2.5 & 1 & 2 & 2.5 & 2.5   &  1041510.251(40)  &  -0.026  & 0.38  \\
			& 2 & 3 & 2.5 & 1.5 & 1 & 2 & 1.5 & 1.5   &  1041510.251(40)  &  -0.026  & 0.62  \\
			& 2 & 3 & 2.5 & 2.5 & 1 & 2 & 1.5 & 2.5   &  1041516.279(40)  &  -0.058  &       \\
			& 2 & 3 & 2.5 & 3.5 & 1 & 2 & 2.5 & 3.5   &  1041550.924(40)  &   0.018  & 0.60  \\
			& 2 & 3 & 2.5 & 1.5 & 1 & 2 & 2.5 & 1.5   &  1041550.924(40)  &   0.018  & 0.15  \\
			& 2 & 3 & 2.5 & 2.5 & 1 & 2 & 2.5 & 2.5   &  1041550.924(40)  &   0.018  & 0.24  \\
			& 2 & 2 & 1.5 & 1.5 & 1 & 1 & 1.5 & 0.5   &  1043854.642(40)  &  -0.025  & 0.07  \\
			& 2 & 2 & 1.5 & 2.5 & 1 & 1 & 1.5 & 1.5   &  1043854.642(40)  &  -0.025  & 0.08  \\
			& 2 & 2 & 1.5 & 0.5 & 1 & 1 & 1.5 & 0.5   &  1043854.642(40)  &  -0.025  & 0.10  \\
			& 2 & 2 & 1.5 & 1.5 & 1 & 1 & 1.5 & 1.5   &  1043854.642(40)  &  -0.025  & 0.18  \\
			& 2 & 2 & 1.5 & 0.5 & 1 & 1 & 1.5 & 1.5   &  1043854.642(40)  &  -0.025  & 0.08  \\
			& 2 & 2 & 1.5 & 2.5 & 1 & 1 & 1.5 & 2.5   &  1043854.642(40)  &  -0.025  & 0.41  \\
			& 2 & 2 & 1.5 & 1.5 & 1 & 1 & 1.5 & 2.5   &  1043854.642(40)  &  -0.025  & 0.08  \\
			& 2 & 2 & 2.5 & 1.5 & 1 & 1 & 1.5 & 0.5   &  1043869.811(40)  &   0.031  & 0.17  \\
			& 2 & 2 & 2.5 & 2.5 & 1 & 1 & 1.5 & 1.5   &  1043869.811(40)  &   0.031  & 0.28  \\
			& 2 & 2 & 2.5 & 1.5 & 1 & 1 & 1.5 & 1.5   &  1043869.811(40)  &   0.031  & 0.05  \\
			& 2 & 2 & 2.5 & 3.5 & 1 & 1 & 1.5 & 2.5   &  1043869.811(40)  &   0.031  & 0.44  \\
			& 2 & 2 & 2.5 & 2.5 & 1 & 1 & 1.5 & 2.5   &  1043869.811(40)  &   0.031  & 0.05  \\
			& 2 & 2 & 2.5 & 1.5 & 1 & 1 & 1.5 & 2.5   &  1043869.811(40)  &   0.031  & 0.00  \\
			& 2 & 2 & 1.5 & 1.5 & 1 & 1 & 0.5 & 0.5   &  1043882.279(40)  &   0.047  & 0.19  \\
			& 2 & 2 & 1.5 & 0.5 & 1 & 1 & 0.5 & 0.5   &  1043882.279(40)  &   0.047  & 0.15  \\
			& 2 & 2 & 1.5 & 2.5 & 1 & 1 & 0.5 & 1.5   &  1043882.279(40)  &   0.047  & 0.50  \\
			& 2 & 2 & 1.5 & 1.5 & 1 & 1 & 0.5 & 1.5   &  1043882.279(40)  &   0.047  & 0.15  \\
			& 2 & 2 & 1.5 & 0.5 & 1 & 1 & 0.5 & 1.5   &  1043882.279(40)  &   0.047  & 0.02  \\
			& 2 & 1 & 0.5 & 1.5 & 1 & 0 & 0.5 & 0.5   &  1063506.127(40)  &   0.013  & 0.32  \\
			& 2 & 1 & 0.5 & 1.5 & 1 & 0 & 0.5 & 1.5   &  1063506.127(40)  &   0.013  & 0.68  \\
			& 2 & 1 & 1.5 & 2.5 & 1 & 0 & 0.5 & 1.5   &  1063578.610(40)  &  -0.036  &       \\
			& 2 & 2 & 2.5 & 3.5 & 1 & 2 & 2.5 & 3.5   &  1067978.220(40)  &   0.058  &       \\
			$v=1$ & 1 & 2 & 2.5 & 3.5 & 0 & 1 & 1.5 & 2.5   &   502775.779(60)  &  -0.063  &       \\
			& 1 & 2 & 2.5 & 2.5 & 0 & 1 & 1.5 & 1.5   &   502777.216(60)  &   0.054  & 0.63  \\
			& 1 & 2 & 2.5 & 1.5 & 0 & 1 & 1.5 & 0.5   &   502777.216(60)  &   0.054  & 0.37  \\
			& 1 & 2 & 1.5 & 2.5 & 0 & 1 & 0.5 & 1.5   &   502780.683(60)  &   0.052  &       \\
			& 1 & 2 & 2.5 & 1.5 & 0 & 1 & 1.5 & 1.5   &   502789.681(60)  &  -0.062  & 0.35  \\
			& 1 & 2 & 1.5 & 0.5 & 0 & 1 & 0.5 & 0.5   &   502789.681(60)  &  -0.062  & 0.65  \\
			& 1 & 2 & 2.5 & 2.5 & 0 & 1 & 1.5 & 2.5   &   502792.545(60)  &   0.031  &       \\
			& 1 & 2 & 1.5 & 1.5 & 0 & 1 & 0.5 & 1.5   &   502798.877(60)  &   0.009  &       \\
			& 1 & 2 & 1.5 & 2.5 & 0 & 1 & 1.5 & 2.5   &   502826.422(60)  &  -0.021  &       \\
			& 1 & 1 & 1.5 & 2.5 & 0 & 1 & 0.5 & 1.5   &   526777.091(60)  &  -0.019  &       \\
			& 1 & 1 & 1.5 & 2.5 & 0 & 1 & 1.5 & 2.5   &   526822.941(60)  &   0.019  &       \\
			\hline\hline \\
	\end{tabular}}
	\begin{minipage}{0.5\textwidth}
		\textbf{Notes.} \\
		Number in parentheses are the experimental uncertainties in units of the last quoted digit.
		The relative weight is given only for blended transitions.
	\end{minipage}
\end{table}

\begin{table*}[!t]
  \caption[]{Summary of the data used for the multi-isotopologue fit of imidogen}
  \label{tab:data}
  \centering\small
  \vspace{2ex}
  \scalebox{0.8}{
  \begin{tabular}{r ccc c ccc}
  \hline \\[-1ex]
           & \mcl{3}{c}{Pure rotational}  &  &  \mcl{3}{c}{Ro-vibrational}  \\
  \cline{2-4} \cline{6-8} \\[-1ex]
           &  no. of lines  &  no of. vib states  & Refs.  &  &  no. of lines & no of. bands  & Refs. \\
  \hline \\[-1ex]
  NH        &   96  &   2  & \citet{Flores2004}, \citet{Lewen2004}, TW                  &  &  451  &   6  & \citet{Bernath1982},\citet{Geller1991}, \citet{Ram2010} \\[0.5ex]
  ND        &  144  &   5  & \citet{Saito1993}, \citet{Takano1998}, \citet{Dore2011}, TW &  &  406  &   6  & \citet{Ram1996}  \\[0.5ex]
  $^{15}$NH &   61  &   2  & \citet{Bailleux2012}, \citet{Bizzocchi2018}                &  &   --  &  --  \\[0.5ex]
  $^{15}$ND &   43  &   2  & This work (TW)                                            &  &   --  &  --  \\[0.5ex]
  \hline
  \end{tabular}}
\end{table*}

\section{Analysis} \label{sec:anal}

\subsection{Effective Hamiltonian} \label{sec:ham}
\indent\indent
From a spectroscopic point of view, imidogen is a free radical with a \elgs ground 
electronic state and exhibits a fine structure due to the dipole-dipole interaction 
of the two unpaired electron spins and to the magnetic coupling of the molecular rotation 
with the total electron spin. 
The couplings of the various angular momenta in NH are described more appropriately by 
Hund's case ($b$) scheme
\begin{equation} \label{eq:coupl1}
 \mathbf{J} = \mathbf{N} + \mathbf{S} \,,
\end{equation}
where \textbf{N} represents the pure rotational angular momentum.
Each fine-structure level is thus labeled by $J, N$ quantum numbers, where 
$J = N + 1, N, N - 1$.
For $N = 0$, only one component ($J = 1$) exists.
Inclusion of the nitrogen and hydrogen hyperfine interactions leads to the couplings
\begin{equation} \label{eq:coupl2}
 \mathbf{F}_1 = \mathbf{J} + \mathbf{I}_\mrm{N} \,, \qquad 
 \mathbf{F} = \mathbf{F}_1 + \mathbf{I}_\mrm{H} \,.
\end{equation}

For each isotopologue in a given ro-vibronic state, the effective Hamiltonian can be 
written as
\begin{equation} \label{eq:Heff}
 \mathbf{H} = \mathbf{H}_\mrm{rv} + \mathbf{H}_\mrm{fs} + \mathbf{H}_\mrm{hfs}
\end{equation} 
where $\mathbf{H}_\mrm{rv}$, $\mathbf{H}_\mrm{fs}$ and $\mathbf{H}_\mrm{hfs}$ are the ro-vibrational, 
fine- and hyperfine-structure Hamiltonians, respectively: 
\begin{equation} \label{eq:Hvr}
 \mathbf{H}_\mrm{rv} = G_v + B_v \mathbf{N}^2 - D_v \mathbf{N}^4  + H_v \mathbf{N}^6 + L_v \mathbf{N}^8 +
                    M_v \mathbf{N}^{10}
\end{equation}
\begin{equation} \label{eq:Hfine}
 \mathbf{H}_\mrm{fs} = \frac{2}{3} \left( \lambda_v + \lambda_{Nv} \mathbf{N}^2\right) \left(3S_z^2 - \mathbf{S}^2\right)
                 + \left(\gamma_v + \gamma_{Nv} \mathbf{N}^2 \right) \mathbf{N}\cdot\mathbf{S}
\end{equation}
\begin{equation} \label{eq:Hhfs}
\begin{split}
 \mathbf{H}_\mrm{hfs} = \sum_i b_{F,v}(i)\,\mathbf{I}_i\cdot\mathbf{S} 
                   + \sum_i c_v(i)\, \left(I_{iz}S_z-\frac{1}{3}\mathbf{I}_i\cdot\mathbf{S}\right) \\
                   + \sum_i eQq_v(i)\,\frac{(3I_{iz}^2-\mathbf{I}_i^2)}{4I_i(2I_i-1)}
                   + \sum_i C_{I,v}(i)\,\mathbf{I}_i\cdot\mathbf{N}
\end{split}
\end{equation}
Here, $G_v$ is the pure vibrational energy, $B_v$ the rotational constant, $D_v$, $H_v$, 
$L_v$, and $M_v$ the centrifugal distortion parameters up to the fifth power in the $\hat{\mathbf{N}}^2$ expansion, 
$\lambda_v$ and $\lambda_{Nv}$ are the electron spin--spin interaction parameter and its 
centrifugal distortion coefficient; $\gamma_v$, and $\gamma_{Nv}$ are the electron spin--rotation 
constant and its centrifugal distortion coefficient, respectively.
The constants $b_{F,v}$ and $c_v$ are the isotropic (Fermi contact interaction) and 
anisotropic parts of the electron spin--nuclear spin coupling, $eQq_v$ represents the 
electric quadrupole interaction and $C_{I,v}$ is the nuclear spin--rotation parameter. 
In Eq.~\eqref{eq:Hhfs} the index $i$ runs over the different nuclei present in a given 
isotopologue. 
The four nuclear spins are: $I$ = 1/2 for H and $^{15}$N and $I$ = 1 for D and $^{14}$N.

\begin{table*}[htb]
  \caption[]{Ro-vibrational Dunham $Y_{lm}$ constants and isotopically invariant $U_{lm}$ 
             parameters determined in the multi-isotopologue fit for imidogen radical}
  \label{tab:rovib}
  \centering\small
  \vspace{2ex}
  \begin{tabular}{lll ld{14} c ld{14}}
    \hline \\[-1ex]
    & & & \mcl{2}{c}{$Y_{lm}$}  & &  \mcl{2}{c}{$U_{lm}$} \\
    \cline{4-5} \cline{7-8} \\[-1ex]
    & $l$ & $m$ &  units & \mcl{1}{c}{value} & & units & \mcl{1}{c}{value} \\[0.5ex]
    \hline \\[-1.5ex]
   &  1  &  0  &  / \wn  &   3282.3629(39)                &  &  \wn u$^{1/2}$   &   3184.2027(35)                   \\ 
   &  2  &  0  &  / \wn  &    -78.6810(46)                &  &  \wn u           &    -73.9831(43)                   \\ 
   &  3  &  0  &  / \wn  &      0.2223(25)                &  &  \wn u$^{3/2}$   &      0.2027(23)                   \\ 
   &  4  &  0  &  / \wn  &     -0.02953(68)               &  &  \wn u$^2$       &     -0.02610(60)                  \\ 
   &  5  &  0  &  / \wn  &     -0.000263(88)              &  &  \wn u$^{5/2}$   &     -0.000225(75)                 \\ 
   &  6  &  0  &  / \wn  &     -0.0001393(45)             &  &  \wn u$^{5/2}$   &     -0.0001158(37)                \\ 
   &  0  &  1  &  / MHz  &      499690.529(84)            &  &  \wn u           &     15.7043731(39)                \\ 
   &  1  &  1  &  / MHz  & -19494.41(34)                  &  &  \wn u$^{3/2}$   &     -0.593919(10)                 \\ 
   &  2  &  1  &  / MHz  &     67.19(48)                  &  &  \wn u$^2$       &      0.001981(14)                 \\ 
   &  3  &  1  &  / MHz  &     -7.69(31)                  &  &  \wn u$^{5/2}$   &     -2.201(87) \times 10^{-4}     \\ 
   &  4  &  1  &  / MHz  &     -1.579(94)                 &  &  \wn u$^3$       &     -4.37(26) \times 10^{-5}      \\ 
   &  5  &  1  &  / MHz  &      0.130(14)                 &  &  \wn u$^{7/2}$   &      3.48(37) \times 10^{-6}      \\ 
   &  6  &  1  &  / MHz  &     -0.01456(76)               &  &  \wn u$^{7/2}$   &     -3.79(20) \times 10^{-7}      \\ 
   &  0  &  2  &  / MHz  &    -51.44722(91)               &  &  \wn u$^2$       &     -0.00152786(15)               \\ 
   &  1  &  2  &  / MHz  &      0.8253(24)                &  &  \wn u$^{5/2}$   &      2.3594(68) \times 10^{-5}    \\ 
   &  2  &  2  &  / MHz  &     -0.0642(15)                &  &  \wn u$^3$       &     -1.779(42) \times 10^{-6}     \\ 
   &  3  &  2  &  / MHz  &      0.00269(37)               &  &  \wn u$^{7/2}$   &      7.24(99) \times 10^{-8}      \\ 
   &  4  &  2  &  / MHz  &     -0.001460(28)              &  &  \wn u$^4$       &     -3.806(72) \times 10^{-8}     \\ 
   &  0  &  3  &  / MHz  &      0.0037950(79)             &  &  \wn u$^3$       &      1.0931(36) \times 10^{-7}    \\ 
   &  1  &  3  &  / MHz  &     -1.339(45) \times 10^{-4}  &  &  \wn u$^{7/2}$   &     -3.60(12) \times 10^{-9}      \\ 
   &  2  &  3  &  / MHz  &     -1.25(18) \times 10^{-5}   &  &  \wn u$^4$       &     -3.26(47) \times 10^{-10}     \\ 
   &  3  &  3  &  / MHz  &     -4.26(31) \times 10^{-6}   &  &  \wn u$^{9/2}$   &     -1.075(79) \times 10^{-10}    \\ 
   &  0  &  4  &  / MHz  &     -4.47(16) \times 10^{-7}   &  &  \wn u$^4$       &     -1.166(42) \times 10^{-11}    \\ 
   &  1  &  4  &  / MHz  &     -3.49(29) \times 10^{-8}   &  &  \wn u$^{9/2}$   &     -8.80(74) \times 10^{-13}     \\    
   &  0  &  5  &  / MHz  &      3.72(88) \times 10^{-11}  &  &  \wn u$^{9/2}$   &      9.1(22) \times 10^{-16}      \\ [0.5ex]    
  \hline \\[-1.5ex]
  \mcl{1}{c}{X}  & $l$ & $m$ & & \mcl{1}{c}{$\delta^\mrm{X}_{lm}$} & & & \mcl{1}{c}{$\Delta^\mrm{X}_{lm}$}  \\[0.5ex]
  \hline \\[-1.5ex]
 N &  0  &  1  &  / MHz  &      75.71(11)              &  &                  &    -3.8592(58)    \\
 N &  1  &  1  &  / MHz  &      -3.45(13)              &  &                  &    -4.50(17)      \\
 H &  1  &  0  &  / \wn  &       1.6117(14)            &  &                  &    -0.90162(78)   \\
 H &  2  &  0  &  / \wn  &      -0.01098(24)           &  &                  &    -0.2565(55)    \\
 H &  0  &  1  &  / MHz  &    1005.124(35)             &  &                  &    -3.68744(13)   \\
 H &  1  &  1  &  / MHz  &      -0.3733(47)            &  &                  &    -3.2030(29)    \\
 H &  0  &  2  &  / MHz  &     -34.054(31)             &  &                  &   -13.23(17)      \\
 H &  0  &  3  &  / MHz  &    1.48(11) \times 10^{-4}  &  &                  &   -69.1(51)       \\
 \hline \\
  \end{tabular}
  \begin{minipage}{0.8\textwidth}
    \textbf{Notes.} \\
    The Dunham constants $Y_{lm}$ are referred to the most abundant NH isotopologue.
    The BOB coefficients $\Delta^\mrm{X}_{lm}$ are adimensional.
    Number in parentheses are the $1\sigma$ statistical errors in units of the last quoted digit.
  \end{minipage}
\end{table*}

\begin{table*}[ht!]
  \caption[]{Fine and hyperfine Dunham $\varY_{lm}$ constants and isotopically invariant 
             $U^y_{lm}$ parameters determined in the multi-isotopologue fit for NH}
  \label{tab:fhf}
  \centering\small
  \vspace{2ex}
  \begin{tabular}{lld{14} c lld{14}}
   \hline \\[-1ex]
   \mcl{3}{c}{Dunham type} & & \mcl{3}{c}{Isotopically invariant} \\[0.5ex]
   \hline \\[-1.5ex]
   \mcl{7}{c}{Fine structure parameters} \\[0.5ex]
   $\lambda_{00}$   & / MHz  &   27573.424(23)              &  &  $U^\lambda_{00}$   & / MHz u         &   0.9174536(52)            \\[0.5ex]
   $\lambda_{10}$   & / MHz  &      16.200(46)              &  &  $U^\lambda_{10}$   & / MHz u$^{1/2}$ &   5.864(20) \times 10^{-4} \\[0.5ex]
   $\lambda_{20}$   & / MHz  &     -14.645(22)              &  &  $U^\lambda_{20}$   & / MHz u         &  -4.5922(68) \times 10^{-4} \\[0.5ex]
   $\lambda_{01}$   & / MHz  &       0.0109(38)             &  &  $U^\lambda_{01}$   & / MHz u$^{3/2}$ &   3.4(12) \times 10^{-7}    \\[0.5ex]
   $\gamma_{00}$    & / MHz  &   -1688.280(31)              &  &  $U^\gamma_{00}$    & / MHz u         &  -0.0528438(13)             \\[0.5ex]
   $\gamma_{10}$    & / MHz  &      88.172(93)              &  &  $U^\gamma_{10}$    & / MHz u$^{3/2}$ &   0.0026750(27)             \\[0.5ex]
   $\gamma_{20}$    & / MHz  &      -1.387(79)              &  &  $U^\gamma_{20}$    & / MHz u$^{3/2}$ &  -4.06(23) \times 10^{-5}   \\[0.5ex]
   $\gamma_{30}$    & / MHz  &       0.370(21)              &  &  $U^\gamma_{30}$    & / MHz u$^{3/2}$ &   1.054(61) \times 10^{-5}  \\[0.5ex]
   $\gamma_{01}$    & / MHz  &       0.4631(31)             &  &  $U^\gamma_{01}$    & / MHz u$^2$     &   1.3656(91) \times 10^{-5} \\[0.5ex]
   $\gamma_{11}$    & / MHz  &      -0.0291(81)             &  &  $U^\gamma_{11}$    & / MHz u$^{3/2}$ &  -8.4(23) \times 10^{-7}    \\[0.5ex]
   $\gamma_{21}$    & / MHz  &       0.0152(45)             &  &  $U^\gamma_{21}$    & / MHz u$^2$     &   4.2(13) \times 10^{-7}    \\[0.5ex]
   $\gamma_{31}$    & / MHz  &      -0.00333(72)            &  &  $U^\gamma_{31}$    & / MHz u$^2$     &  -9.0(19) \times 10^{-8}    \\[0.5ex]
   $\delta^{\lambda,\mrm{N}}_{00}$ & / MHz &   -3.59(16)    &  &  $\Delta^{\lambda,\mrm{N}}_{00}$    & &     0.837(37)              \\[0.5ex]
   $\delta^{\lambda,\mrm{H}}_{00}$ & / MHz &  -65.250(40)   &  &  $\Delta^{\lambda,\mrm{H}}_{00}$    & &     1.08957(67)            \\[0.5ex]
   $\delta^{\lambda,\mrm{H}}_{10}$ & / MHz &    1.934(40)   &  &  $\Delta^{\lambda,\mrm{H}}_{10}$    & &   -49.0(10)                \\[0.5ex]
   $\delta^{\gamma,\mrm{H}}_{00}$  & / MHz &    1.712(18)   &  &  $\Delta^{\gamma,\mrm{H}}_{00}$     & &     3.515(38)              \\[1.5ex]
   $\delta^{\gamma,\mrm{H}}_{10}$  & / MHz &   -0.091(17)   &  &  $\Delta^{\gamma,\mrm{H}}_{10}$     & &     3.83(66)               \\[1.5ex]
   \mcl{7}{c}{Hyperfine structure parameters} \\[0.5ex]
   $b_{F,00}(\mrm{H})$       & \ MHz & -64.194(13)  & &  $U^{b_F}_{00}(\mrm{H})$     & \wn           &  -0.00214129(45)            \\[0.5ex]
   $b_{F,10}(\mrm{H})$       & \ MHz &  -3.785(15)  & &  $U^{b_F}_{10}(\mrm{H})$     & \wn           &  -1.2243(48) \times 10^{-4} \\[0.5ex]
   $c_{00}(\mrm{H})$         & \ MHz &  92.216(61)  & &  $U^{c}_{00}(\mrm{H})$       & \wn           &   0.0030760(20)             \\[0.5ex]
   $c_{10}(\mrm{H})$         & \ MHz &  -3.391(58)  & &  $U^{c}_{10}(\mrm{H})$       & \wn           &  -1.097(19) \times 10^{-4}  \\[0.5ex]
   $C_{00}(\mrm{H})$         & \ MHz &  -0.068(11)  & &  $U^{C}_{00}(\mrm{H})$       & \wn           &  -2.14(33) \times 10^{-6}   \\[0.5ex]
   $b_{F,00}(\mrm{D})$       & \ MHz &  -9.836(12)  & &  $U^{b_F}_{00}(\mrm{D})$     & \wn           &  -3.2810(41) \times 10^{-4} \\[0.5ex]
   $b_{F,10}(\mrm{D})$       & \ MHz &  -0.632(10)  & &  $U^{b_F}_{10}(\mrm{D})$     & \wn u$^{1/2}$ &  -2.044(34) \times 10^{-5}  \\[0.5ex] 
   $c_{00}(\mrm{D})$         & \ MHz &  14.233(74)  & &  $U^{c}_{00}(\mrm{D})$       & \wn           &   4.746(25) \times 10^{-4}  \\[0.5ex]
   $c_{10}(\mrm{D})$         & \ MHz &  -0.372(99)  & &  $U^{c}_{10}(\mrm{D})$       & \wn u$^{1/2}$ &  -1.19(32) \times 10^{-5}   \\[0.5ex]
   $c_{20}(\mrm{D})$         & \ MHz &  -0.117(32)  & &  $U^{c}_{20}(\mrm{D})$       & \wn u$^{1/2}$ &  -3.66(100) \times 10^{-6}  \\[0.5ex]
   $eQe_{00}(\mrm{D})$       & \ MHz &   0.080(32)  & &  $U^{eQq}_{00}(\mrm{D})$     & \wn           &   2.7(11) \times 10^{-6}    \\[0.5ex]
   $b_{F,00}(^{14}\mrm{N}$)  & \ MHz &  19.084(18)  & & $U^{b_F}_{00}(^{14}\mrm{N}$) & \wn           &   6.3660(59) \times 10^{-4} \\[0.5ex]
   $b_{F,10}(^{14}\mrm{N}$)  & \ MHz &  -0.421(31)  & & $U^{b_F}_{10}(^{14}\mrm{N}$) & \wn u$^{1/2}$ &  -1.365(99) \times 10^{-5}  \\[0.5ex]
   $b_{F,20}(^{14}\mrm{N}$)  & \ MHz &  -0.088(13)  & & $U^{b_F}_{20}(^{14}\mrm{N}$) & \wn u$^{1/2}$ &  -2.76(41) \times 10^{-6}   \\[0.5ex]
   $c_{00}(^{14}\mrm{N}$)    & \ MHz & -68.135(31)  & & $U^{c}_{00}(^{14}\mrm{N}$)   & \wn u$^{1/2}$ &  -0.0022727(10)             \\[0.5ex]
   $c_{10}(^{14}\mrm{N}$)    & \ MHz &   0.467(26)  & & $U^{c}_{10}(^{14}\mrm{N}$)   & \wn u$^{3/2}$ &   1.509(83) \times 10^{-5}  \\[0.5ex]
   $eQq_{00}(^{14}\mrm{N}$)  & \ MHz &  -3.367(54)  & & $U^{eQq}_{00}(^{14}\mrm{N}$) & \wn           &  -1.123(18) \times 10^{-4}  \\[0.5ex]
   $eQq_{10}(^{14}\mrm{N}$)  & \ MHz &   0.395(38)  & & $U^{eQq}_{10}(^{14}\mrm{N}$) & \wn u$^{1/2}$ &   1.28(12) \times 10^{-5}   \\[0.5ex]
   $C_{00}(^{14}\mrm{N}$)    & \ MHz &   0.172(11)  & & $U^{C}_{00}(^{14}\mrm{N}$)   & \wn u$^{1/2}$ &   5.42(36) \times 10^{-6}   \\[0.5ex]
   $C_{10}(^{14}\mrm{N}$)    & \ kHz &  -0.0293(85) & & $U^{C}_{10}(^{14}\mrm{N}$)   & \wn u$^{3/2}$ &  -9.0(26) \times 10^{-7}    \\[0.5ex]
   $b_{F,00}(^{15}\mrm{N}$)  & \ MHz & -26.848(16)  & & $U^{b_F}_{00}(^{15}\mrm{N}$) & \wn           &  -8.9556(52) \times 10^{-4} \\[0.5ex]
   $b_{F,10}(^{15}\mrm{N}$)  & \ MHz &   0.860(18)  & & $U^{b_F}_{10}(^{15}\mrm{N}$) & \wn u$^{1/2}$ &   2.789(58) \times 10^{-5}  \\[0.5ex]
   $c_{00}(^{15}\mrm{N}$)    & \ MHz &  95.428(49)  & & $U^{c}_{00}(^{15}\mrm{N}$)   & \wn           &   0.0031831(16)             \\[0.5ex]
   $c_{10}(^{15}\mrm{N}$)    & \ MHz &  -0.556(54)  & & $U^{c}_{10}(^{15}\mrm{N}$)   & \wn           &  -1.80(17) \times 10^{-5}   \\[0.5ex]
   $C_{00}(^{15}\mrm{N}$)    & \ MHz &  -0.259(22)  & & $U^{C}_{00}(^{15}\mrm{N}$)   & \wn u$^{1/2}$ &  -8.18(71) \times 10^{-6}   \\[0.5ex]
   $C_{10}(^{15}\mrm{N}$)    & \ MHz &   0.099(24)  & & $U^{C}_{10}(^{15}\mrm{N}$)   & \wn u$^{1/2}$ &   3.06(74) \times 10^{-6}   \\[0.5ex]
   \hline \\
  \end{tabular}
  \begin{minipage}{0.8\textwidth}
    \textbf{Notes.} \\
    The Dunham constants $\varY_{lm}$ are referred to the most abundant NH isotopologue.
    The BOB coefficients $\Delta^\mrm{X}_{lm}$ are adimensional.
    Number in parentheses are the $1\sigma$ statistical errors in units of the last quoted digit.
  \end{minipage}
\end{table*}

\subsection{Multi-isotopologue Dunham models} \label{sec:duhn}
\indent\indent
In order to treat the data of all the available isotopologues in a global analysis,
it is convenient to adopt a Dunham-type expansion \cite{Dunham1932}.
The ro-vibrational energy levels are given by the equation:
\begin{equation} \label{eq:rvDunham}
 E_\mrm{rv}(v,N) = \sum_{l,m} Y_{lm} \left(\varv + \tfrac{1}{2}\right)^l [N(N + 1)]^m \,.
\end{equation}
The fine- and hyperfine-structure parameters [i.e., $\lambda_v$, $\lambda_{Nv}$, $\gamma_v$, 
$\gamma_{Nv}$,  $b_{F,v}$, $c_v$, $eQq_v$, and $ C_{I,v}$ in 
Eqs.~\eqref{eq:Hfine}--\eqref{eq:Hhfs}], are given by analogous expansions:
\begin{equation} \label{eq:XDunh}
 y(v,N) = \sum_{l,m} \varY_{lm} \left(v + \tfrac{1}{2}\right)^l [N(N + 1)]^m \,,
\end{equation}
where $y(v,N)$ represents the effective value of the parameter $y$ in the ro-vibrational level labeled by $(v,N)$, and $\varY_{lm}$ are the coefficients of its 
Dunham-type expansion
\footnote{%
  This ($v,N$)-factorisation is possible because all the angular momentum 
  operators multiplying the coefficients of Eqs.~\eqref{eq:Hfine} and~\eqref{eq:Hhfs} 
  commute with purely vibrational operators and with $\mathbf{N}^2$\@.}\@. \\
The spectroscopic constants of Eqs.~\eqref{eq:Hvr}--\eqref{eq:Hhfs} can be expressed in 
terms of the Dunham coefficients $Y_{lm}$ and $\varY_{lm}$\@.
For example, the constants $G_v$, $B_v$, and $D_v$, of the ro-vibrational Hamiltonian are given 
by the following expansions: 
\begin{subequations} \label{eq:rvexp}
\begin{align}
 G_v &= \sum_{l=0} Y_{l0} \left(v + \tfrac{1}{2}\right)^l \,, \\
 B_v &= \sum_{l=0} Y_{l1} \left(v + \tfrac{1}{2}\right)^l \,, \\
 D_v &= \sum_{l=0} Y_{l2} \left(v + \tfrac{1}{2}\right)^l \,.
\end{align}
\end{subequations}

Each fine- and hyperfine-structure constant is also expressed by suitable expansions.
For example, the electron spin-rotation constant and its centrifugal dependence in a given vibrational state $ \varv $ can be expressed as:
\begin{subequations} \label{eq:fsexp}
\begin{align}
 \gamma_{\,v}    &= \sum_{l=0} \gamma_{\,l0} \left(v + \tfrac{1}{2}\right)^l \,, \\
 \gamma_{\,Nv} &= \sum_{l=0} \gamma_{\,l1} \left(v + \tfrac{1}{2}\right)^l \,,
\end{align}
\end{subequations}

where $ \gamma_{\,l0} $ and $ \gamma_{\,l1} $ are the $ \varY_{lm} $ constants of Eq.~\ref{eq:XDunh} relative to the spin-rotation interaction.
For a given isotopologue $\alpha$, a specific set of Dunham constants $Y^{(\alpha)}_{lm}$ 
and $\varY^{(\alpha)}_{lm}$ is defined.
Such constants can be described in terms of isotopically invariant parameters
using the known reduced mass dependences given by \cite{Ross1974,Watson1980}
\begin{subequations} \label{eq:massscW}
\begin{align}
 Y^{(\alpha)}_{lm} &= U_{lm} \mu_\alpha^{-(l/2 + m)} 
            \left[1 + m_e\left(\frac{\Delta^\mrm{N}_{lm}}{M^{(\alpha)}_\mrm{N}} + 
                               \frac{\Delta^\mrm{H}_{lm}}{M^{(\alpha)}_\mrm{H}}\right)\right] \,, 
                               \label{eq:masssc1} \\
 \mathpzc{Y}^{(\alpha)}_{lm} &= U^y_{lm} \mu_\alpha^{-(l/2 + m + p)} 
            \left[1 + m_e\left(\frac{\Delta^{y,\mrm{N}}_{lm}}{M^{(\alpha)}_\mrm{N}} +
                                \frac{\Delta^{y,\mrm{H}}_{lm}}{M^{(\alpha)}_\mrm{H}}\right)\right] \,, 
                                \label{eq:masssc2}
\end{align}
\end{subequations}
where $M^{(\alpha)}_\mrm{X}$ (with $X=\mrm{N},\mrm{H}$) are the atomic masses, 
$\mu_\alpha$ is the reduced mass of the $\alpha$ isotopologue, and $m_e$ is the electron 
mass. 
$U_{lm}$ and $U^y_{lm}$ are isotopically invariant Dunham constants, whereas 
$\Delta^\mrm{X}_{lm}$ and $\Delta^{y,\mrm{X}}_{lm}$ are unitless coefficients which 
account for the Born--Oppenheimer Breakdown \cite{Watson1973,Watson1980}.
In Eq.~\eqref{eq:masssc2}, $p=0$ for $y = \lambda, b_F, c, eQq$, while $p=1$ for 
$y = \gamma, C_I$.
This extra $\mu^{-1}$ factor in the mass scaling is needed to account for the 
intrinsic $\mathbf{N}^2$ dependence of the spin--rotation constants \cite{Brown1977}.
Here, the unknowns are the $U_{lm}$, $U^y_{lm}$ coefficients and the corresponding 
$\Delta^\mrm{X}_{lm}$ and $\Delta^{y,\mrm{X}}_{lm}$ BOB corrections.

An alternative approach has been proposed by LeRoy \cite{LeRoy1999}, where one isotopologue 
(usually the most abundant one) is chosen as reference species ($\alpha=1$), and the 
Dunham parameters $Y^{(\alpha)}_{lm}$ and $\mathpzc{Y}^{(\alpha)}_{lm}$ of any other 
species are obtained by the following mass scaling
\begin{subequations} \label{eq:massscLR}
\begin{align}
 Y^{(\alpha)}_{lm} &= \left[Y^{(1)}_{lm} + \frac{\Delta M_\mrm{N}}{M^{(\alpha)}_\mrm{N}}\delta^\mrm{N}_{lm} 
                            + \frac{\Delta M_\mrm{H}}{M^{(\alpha)}_\mrm{H}}\delta^\mrm{H}_{lm}
                      \right] \left(\frac{\mu_1}{\mu_\alpha}\right)^{(l/2 + m)} \,,
                      \label{eq:masssc3} \\
 \mathpzc{Y}^{(\alpha)}_{lm} &= \left[\varY^{(1)}_{lm} 
                                + \frac{\Delta M_\mrm{N}}{M^{(\alpha)}_\mrm{N}}\delta^{y,\mrm{N}}_{lm} 
                                + \frac{\Delta M_\mrm{H}}{M^{(\alpha)}_\mrm{H}}\delta^{y,\mrm{H}}_{lm}
                 \right] \left(\frac{\mu_1}{\mu_\alpha}\right)^{(l/2 + m + p)} \,. \label{eq:masssc4}
\end{align}
\end{subequations}
Here, $\Delta M_\mrm{X} = M^{(\alpha)}_\mrm{X} - M^{(1)}_\mrm{X}$ (with $X=\mrm{N},\mrm{H}$) 
are the mass differences produced by the isotopic substitution, with respect to the reference 
species, and the BOB corrections are described by the new $\delta^\mrm{X}_{lm}$ and
$\delta^\mrm{y, X}_{lm}$ coefficients.
These are related to the dimensionless $\Delta^\mrm{X}_{lm}$ of Eqs.~\ref{eq:massscW} through the simple relation
\begin{equation} \label{eq:deltas}
 \Delta^\mrm{X}_{lm} = \delta^\mrm{X}_{lm}\frac{M^{(1)}_X}{m_e}
                       \left(Y^{(1)}_{lm} + \delta^\mrm{N}_{lm} + \delta^\mrm{H}_{lm}\right)^{-1} \,. 
\end{equation}
Albeit formally equivalent, this latter parametrisation was introduced to overcome a 
number of deficiencies of the traditional treatment which were pointed out by Watson 
\cite{Watson1980} and Tiemann \cite{Tiemann1982}, and its features are discussed in great 
detail in the original paper \cite{LeRoy1999}.
An obvious advantage of the alternative mass scaling of Eqs.~\eqref{eq:massscLR} is that 
the fitted coefficients are all expressed in frequency units and are directly linked to the familiar spectroscopic parameters of the reference 
isotopologue (e.g., $Y^{(1)}_{10}\approx\omega_e$, $Y^{(1)}_{20}\approx -\omega_ex_e$, 
$Y^{(1)}_{01}\approx B_e$, $Y^{(1)}_{11}\approx -\alpha_e$, etc.)\@.
Furthermore, the BOB contributions are accounted for using purely addictive terms thus
reducing the correlations among the parameters.

\section{Results} \label{sec:results}

\subsection{\qnd spectrum} \label{sec:res15nd}
\begin{figure}[h!]
\centering
\includegraphics[width=0.47\textwidth]{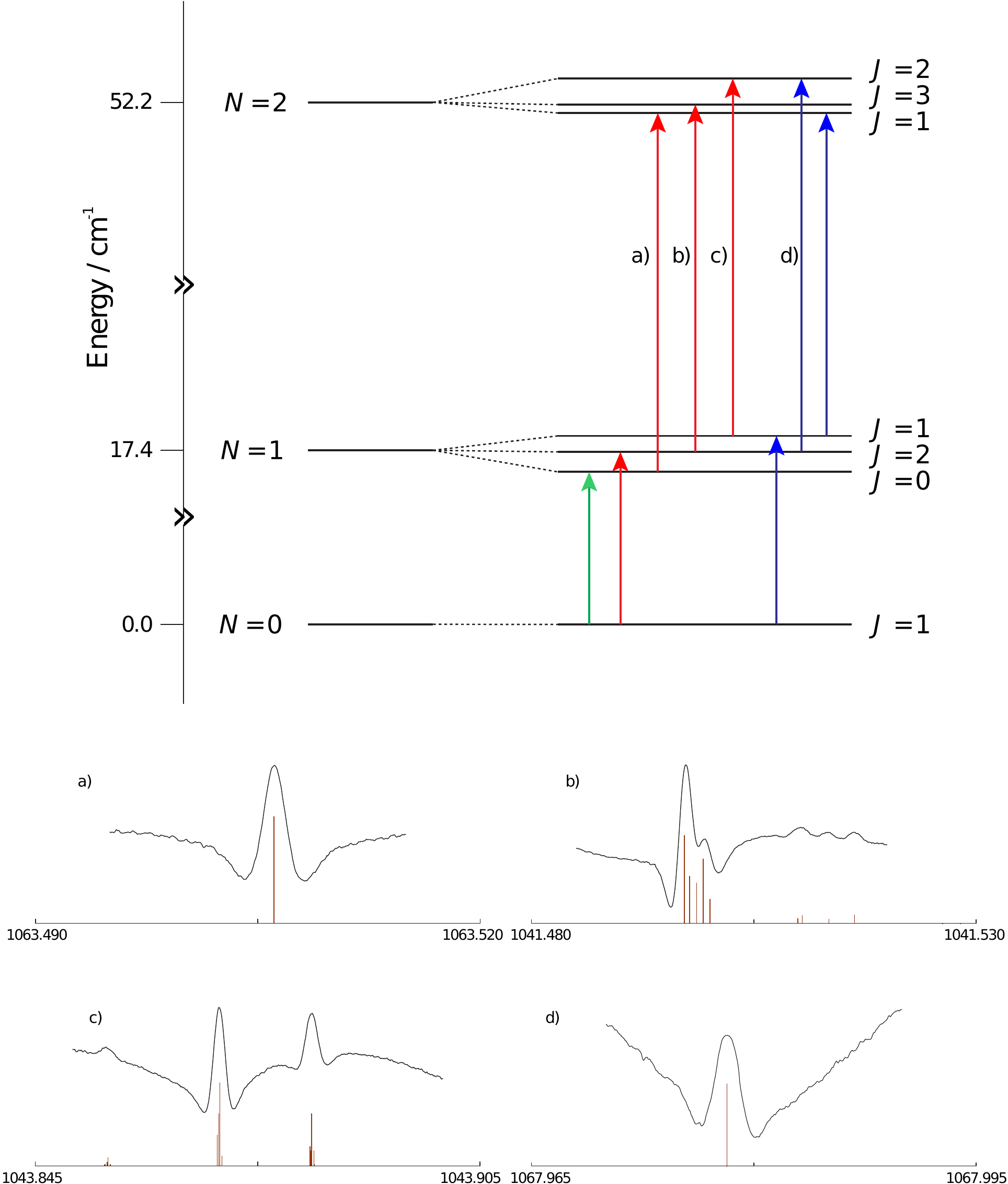}
\caption{\textit{Upper panel}: energy levels scheme of \qnd in the ground vibrational state.
 		  The hyperfine-structure is not shown.
          The arrows mark the transitions observed in this study:
          $\Delta J = +1$ (red),  $\Delta J = 0$ (blue), and $\Delta J = -1$ (green).
          \textit{Lower panel}: spectral recordings for the transitions marked with the labels
          \textit{a}, \textit{b}, \textit{c} and \textit{d} showing the corresponding hyperfine
          structure.
          The brown sticks represent the positions and the intensities of the hyperfine components
          computed from the spectroscopic parameters of Table~\ref{tab:15ndres}.
          \label{fig:livelli}}
\end{figure}

\indent\indent
For the previously unobserved \qnd species, we have recorded 34~lines for the ground 
vibrational state and 9~lines for the $v=1$ state.
They include the complete fine-structure of the $N = 1 \leftarrow 0$ transition and the strongest
fine-components of the $N = 2 \leftarrow 1$ transition for the ground state
(see Figure~\ref{fig:livelli}), and the $\Delta J = 0, + 1$ components of the $N = 1 \leftarrow 0$
transition for the $v=1$ state.
The corresponding transition frequencies were fitted to the Hamiltonian of
Eqs.~\eqref{eq:Hvr}--\eqref{eq:Hhfs} using the \textsc{SPFIT} analysis
program \cite{Pickett1991}.
Because of the small number of transitions detected for the $v=1$ state, some of the 
spectroscopic parameters for this state could not be directly determined in the least-squares
fit and were constrained to the corresponding ground state values.
The two sets of constants for $v=0$ and $v=1$ states are reported in 
Table~\ref{tab:15ndres}.
The list of observed frequencies, along with the residuals from the single-species fit, is given
in Table~\ref{tab:15ndfit}. 
In addition, the .LIN and .PAR input files for the \textsc{SPFIT} programm are
included in the supplementary material \dag.

\subsection{Multi-isotopologue Dunham fit}
\indent\indent
In this work, we carried out a multi-isotopologue Dunham fit of the imidogen radical in its 
\elgs ground electronic state using our newly measured transition frequencies for the 
doubly substituted \qnd variant plus all the available rotational and ro-vibrational data for the NH,
\qnh and ND species.
To take into account the different experimental precision, each datum was given a weight
inversely proportional to the square of its estimated measurement error, $w = 1/\sigma^2$\@.
The $\sigma$ values adopted for the present measurements have been discussed in 
\S~\ref{sec:exp}~, while for literature data, we retained the values provided in each 
original work.

The content of the data set and the relevant bibliographic references are summarised 
in Table~\ref{tab:data}.
In total, the data set contains 1563 ro-vibrational transitions which correspond to 1201 
distinct frequencies.
These data were fitted to the multi-isotopologue model described in 
\S\S~\ref{sec:ham}--\ref{sec:duhn}, using both traditional [Eqs.~\eqref{eq:massscW}] and 
LeRoy [Eqs.~\eqref{eq:massscLR}] mass scaling schemes to describe the Dunham-type 
parameters ($Y_{lm}$ and $\varY_{lm}$) of each isotopic species.

The analysis was performed using a custom \textsc{Python} code which uses the
\textsc{SPFIT} program \cite{Pickett1991} as computational core.
Briefly, the scripting routine reads the atomic masses, the spin 
multiplicities, and the $Y_{lm}$ constants for the reference species.
Then, the \textsc{SPFIT} parameter file (.PAR) is set up by defining several 
sets of spectroscopic constants (one for each isotopologue/vibrational 
state), taking into account the mass scaling factors.
The \textsc{SPFIT} lines file (.LIN) is created by colletting the experimental data.
In this process, half integer quantum numbers are rounded up and a
"quantum number state" is assigned to each isotopologue in a given vibrational state,
in conformity with the \textsc{SPFIT} format.
At the end of the least-squares optimisation, the \textsc{SPFIT} output is 
post-processed, and the final parameters list is reformatted in the
same fashion of the initial input data set.
The atomic masses used were taken from the \citeauthor{Wang2017}
\cite{Wang2017} compilation.
The optimised parameters are reported in Tables~\ref{tab:rovib} and \ref{tab:fhf}, while
the complete list of all the fitted data, together with the residuals from the multi-isotopologue analysis,
is provided as supplementary material (the .LIN and .PAR files are also provided) \dag.

\section{Discussion}

\subsection{Spectroscopic parameters} \label{sec:disc:par}
\indent\indent
From the multi-isotopologue analysis we obtained a highly satisfactory fit.
Its quality can be evaluated in several ways.
First of all, we were able to reproduce the input data within their estimated uncertainties:
the overall standard deviation of the weighted fit is $\sigma = 0.89$, and the root-mean-square deviations of the
residuals computed separately for the rotational and ro-vibrational data are of the same order
of magnitude of the corresponding measurements error, RMS$_\text{ROT}$ = 0.107\,MHz and
RMS$_\text{VIBROT}$ = 3.4$\times$ 10$^{-3}$\,\wn, respectively.
Then, the various sets of $Y_{lm}$ for a given $m$ constitute a series
whose coefficients decrease in magnitude for increasing values of the index $l$, as expected for a
rapidly converging Dunham-type expansion.
In general, most of the determined coefficients have a relative error lower than $5$\%.
Higher errors are observed only for those constants with high $l$-index and this is due to the
smaller number of transitions available for highly vibrationally excited states.
Finally, the Kratzer \cite{Kratz1920} and Pekeris \cite{Pekeris1934} relation can also be used as
a yardstick to asses the correct treatment of the Born--Oppenheimer Breakdown effects.
Using the formula \cite{Gordy1984}
\begin{equation} \label{eq:kratzer}
 Y_{02} \simeq \dfrac{4Y_{01}^3}{Y_{10}^2} \,,
\end{equation}
we obtained for $Y_{02}$ a value of 51.54051\,MHz which compares well with the fitted
one of 51.44722(91)\,MHz\@.

\subsection{Equilibrium bond distance}
\indent\indent
The precision yielded by the high-resolution spectroscopic technique led to a very accurate 
determination of the equilibrium bond length $r_e$ for the imidogen radical.
The rotational measurements of a diatomic molecule in its ground vibrational state ($v = 0$) 
allow the determination of precise value of $r_0$, which includes the zero point vibrational 
contributions and differs from $r_e$.
This latter is determinable from the rotational spectrum in at least one 
vibrationally excited state.
In the present analysis, data of four isotopic species in several vibrational excited states
have been combined, allowing for a very precise determination of $r_e$ for each isotopologue
$\alpha$.
The equilibrium bond distance is given by:
\begin{equation}\label{eq:bondlenght}
 r_e^{(\alpha)} = \sqrt{\dfrac{N_a h}{8\pi^2} \dfrac{1}{B_e^{\left( \alpha \right) }\mu_a}} \,.
\end{equation}
where $ N_ah $ is the molar Planck constant.
Actually, the values of $B^{(\alpha)}_e$ differ from those of $Y^{(\alpha)}_{01}$ obtained 
from the Dunham-type analysis. 
This discrepancy should be ascribed to a small contribution, expressed by \citep{Gordy1984}:
\begin{equation} \label{eq:corrDun}
 Y_{01} = B_e + \Delta Y_{01}^{\left( \text{Dunh} \right)} 
        = B_e + \beta_{01}\left( \frac{B_e^3}{\omega_e^2} \right) \,,
\end{equation} 
with
\begin{equation}
 \beta_{01} = Y^2_{10} \left( \frac{Y_{21}}{4Y^3_{01}} \right)  
            + 16a_1 \left( \frac{Y_{20}}{3 Y_{01}} \right) - 8 a_1 - 6a^2_1 + 4 a^3_1 \,,
\end{equation}
and 
\begin{equation} \label{eq:a1}
 a_1 = \frac{Y_{11}}{3\sqrt{-Y_{02}Y_{01}}} - 1 \,.
\end{equation}
From Eq.~\eqref{eq:bondlenght}, it is evident that the bond length $r_e$ assumes different 
values for each isotopologue. 
On the contrary, by substituting the product $B_e^{(\alpha)}\mu_a$ with $U_{01}$, one obtains 
an isotopically independent equilibrium bond length $r^\mrm{BO}_e$. 
In the present case, $r^\mrm{BO}_e$ takes the value of 103.606721(13)\,pm.
In Table~\ref{tab:bondlenghts}, this result is compared with the
equilibrium bond distances calculated from the $ B_e $ of each isotopologue NH, \qnh, ND, and \qnd.
In this case, $ B_e $ was obtained by correcting the corresponding $ Y_{01} $ constant according to Eqs..~\eqref{eq:corrDun}--\eqref{eq:a1}.  
It should be noticed that the values differ at sub-picometre level but these differences, 
even if small, are detectable thanks to the high-precision of rotational measurements.

The experimental value derived for $r^\mrm{BO}_e$ has been compared with a theoretically
best estimate obtained following the prescriptions of Refs.~\citenum{Heckert2006,Puzz2008}.
A composite calculation have been carried out considering basis-set extrapolation, 
core-correlation effects, and inclusion of higher-order corrections due to the use of the 
full coupled-cluster singles and doubles, augmented by a perturbative treatment of triple
excitation [CCSD(T)] model
\begin{equation}
 \text{fc-CCSD(T)/cc-pV$\infty$Z + $\Delta$core/cc-pCV5Z + $\Delta$T/cc-pVTZ} \,. \notag
\end{equation}
The computation have been performed using CFOUR \cite{CFOUR}.
From this theoretical procedure we obtained $r^\text{\,theor}_e = 103.5915$\,pm (see also Table~\ref{tab:bondlenghts}), which is in very good agreement with the experimentally 
derived value, the discrepancy being 15\,fm\@.

\begin{table}[!t]
 \caption[]{Born--Oppenheimer and equilibrium bond distances (in pm) 
            from the individual isotopologues (see text).
            \label{tab:bondlenghts}}
 \centering\small
 \vspace{2ex}
 \begin{tabular}{r c r}
  \hline \\[-1ex]
  Species    & $r_e$          &  $r_e - r^\mrm{BO}_e$  \\[0.5ex]
  \hline \\[-1.5ex]
  NH         & 103.716377(16) & 0.109656  \\
  \qnh       & 103.715864(16) & 0.109143  \\
  ND         & 103.665420(10) & 0.058699  \\
  \qnd       & 103.664908(10) & 0.058187  \\
  \hline \\[-1.5ex]
  \mcl{3}{c}{$r^\mrm{\,BO}_e = 103.606721(13)$} \\
  \mcl{3}{c}{$r^\text{\,theor}_e = 103.5915$ \pha{111}}   \\[0.5ex]
  \hline 
 \end{tabular} 
\end{table}

\subsection{Born--Oppenheimer Breakdown} \label{sec:disc:adiab}
\indent\indent
The BOB coefficients $\Delta^X_{lm}$  determined in the present analysis account for the 
small inaccuracies of the Born--Oppenheimer approximation in describing the ro-vibrational 
energies of the imidogen radical.
For the rotational constant ($\approx Y_{01}$), it is possible to identify three different 
contributions to the corresponding BOB parameter \cite{Bizzocchi2015}

\begin{equation} \label{eq:BOBt}
\begin{split}
 \Delta^X_{01} = \left(\Delta^X_{01}\right)^\mrm{ad}  + \left(\Delta^{X}_{01}\right)^\mrm{nad} 
               + \left(\Delta^X_{01}\right)^\text{Dunh} \\
               = \left(\Delta^X_{01}\right)^\mrm{ad} 
               + \frac{\left(\mu g_J \right)_{X'}}{m_p}
               + \frac{\Delta Y^{(\text{Dunh})}_{01} \mu}{m_e B_e} \,, 
\end{split}
\end{equation}

namely an adiabatic contribution, a non-adiabatic term, and a Dunham correction, respectively.
The last two terms on the right side of Eq.~\eqref{eq:BOBt} can be computed from purely 
experimental quantities: $\left(\Delta^X_{01}\right)^\text{Dunh}$ arises from the use of a Dunham 
expansion and contains the term $\Delta Y_{01}^{\left( \text{Dunh} \right)}$ of Eq.~\eqref{eq:corrDun}, 
whereas $\left(\Delta^{X}_{01}\right)^\mrm{nad}$ depends on the mixing of the electronic 
ground state with nearby electronic excited states, and can be estimated from the molecular 
electric dipole moment $\mu$ and the rotational $g_J$ factors \cite{Watson1973}.
The adiabatic term can be simply computed as the difference between the experimental $\Delta^X_{01}$
and the terms $\left(\Delta^X_{01}\right)^\text{Dunh}$ and $\left(\Delta^{X}_{01}\right)^\mrm{nad}$.

\citet{Tiemann1982} found that the adiabatic term $\left(\Delta^X_{01}\right)^\mrm{ad}$
basically depends on the corresponding $X$ atom rather than on the 
particular molecular species.
Hence, it is interesting to derive this contribution in order to compare the results 
obtained for different molecules and to verify the reliability of the empirical fitting
procedure.

All the contributions of Eq.~\eqref{eq:BOBt} are collected in Table~\ref{tab:bobterms}.
The non-adiabatic contribution has been computed using the literature value of the dipole
moment\cite{Scarl1974} $\mu = 1.389$\,D  and the ground state $g_J$ value estimated from
a laser magnetic resonance study\cite{Robin2007}, $g_J = 0.001524$\@. 

From the adiabatic contribution to the Born--Oppenheimer Breakdown coefficients for the 
rotational constants, $\left(\Delta^X_{01}\right)^\mrm{ad}$, one may derive the corresponding
correction to the equilibrium bond distance, a quantity which can also be accessed by 
\textit{ab initio} computations.
From our Eq.~\eqref{eq:masssc1} and Eq.~(6) of Ref.~\citenum{Gauss2010}, the following 
equality is obtained
\begin{equation} \label{eq:DRad}
 \Delta R_\mrm{ad} = -\frac{r_e}{2}\left[\frac{m_e}{M_\mrm{N}}\left(\Delta^\mrm{N}_{01}\right)^\mrm{ad}
                                         + \frac{m_e}{M_\mrm{H}}\left(\Delta^\mrm{H}_{01}\right)^\mrm{ad} \right] \,.
\end{equation}
The adiabatic correction to the equilibrium bond distance, $\Delta R_\mrm{ad}$, can be theoretically estimated through the 
computation of the adiabatic bond distance, i.e., the minimum of the potential given 
by the sum of the Born--Oppenheimer potential augmented by the diagonal Born--Oppenheimer 
corrections (DBOC)\cite{Gauss2010}.
The difference between the equilibrium bond distances calculated with and without DBOC, with tight convergence limits, performed 
at the CCSD/cc-pCV$n$Z level ($n = 3,4,5$), yielded $\Delta R_\mrm{ad} = 0.026$\,pm\@.
This value is in very good agreement with the purely experimental one obtained by 
Eq.~\eqref{eq:DRad} which results $0.020$\,pm\@, thus providing a confirmation for the 
validity of our data treatment.

\begin{table}[!t]
 \caption[]{Contributions of the Born--Oppenheimer Breakdown coefficients to the $U_{01}$ constant
            \label{tab:bobterms}}
 \centering\small
 \vspace{2ex}
 \begin{tabular}{r cccc}
  \hline \\[-1ex]
  Atom  & $\Delta_{01}$(exp) &  adiabatic  & non-adiabatic & Dunham  \\[0.5ex]
  \hline \\[-1.5ex]
  N         & -3.8592  & -0.6515   &  -3.1326  &  -0.0751  \\
  H         & -3.6874  & -1.0379   &  -2.5744  &  -0.0751  \\[0.5ex]
  \hline 
 \end{tabular}
\end{table}

\subsection{Zero-point Energy}
\indent\indent
The results of our analysis make possible to estimate the zero-point energy (ZPE) 
for each isotopologue from the Dunham's constants $Y_{lm}$ with $m = 0$, namely: 
\begin{equation}
 G(0) = \sum_{l=0} Y_{l0} \left(\frac{1}{2}\right)^l \,.
\end{equation}
As we determined anharmonicity constants up to the sixth order, the ZPE is derived with a negligible truncation bias \cite{Irikura2007} from the expression:
\begin{equation} \label{eq:zpe}
  G(0) = Y_{00} + \frac{Y_{10}}{2} - \frac{Y_{20}}{4} + \frac{Y_{30}}{8} + \frac{Y_{40}}{16}  
                 + \frac{Y_{50}}{32} + \frac{Y_{60}}{64} \,.
\end{equation}
The $Y_{00}$ constant present in the Dunham-type expansions is not experimentally accessible.
Its value can be estimated, to a good approximation, through\cite{Irikura2007}
\begin{equation}
  Y_{00} \simeq \frac{B_e}{4} + \frac{\alpha_e \omega_e}{12 B_e} + \frac{\alpha^2_e \omega^2_e}{12^2 B^3_e} 
                              - \frac{\omega_ex_e}{4} \,.
\end{equation}
The value for the main isotopologue NH is 1.9987(12) \wn. \\
The values obtained for the ZPE of the four isotopologues are collected in Table~\ref{tab:zpe}. 
For comparison, the values of literature are also reported. 
Our results agree well with those reported in the literature\cite{Irikura2007}, but our 
precision is more than one order of magnitude higher.
The errors on our ZPE values are \textit{ca.} 1 $\times 10^{-3}$ \wn and were calculated taking into account the error propagation 

\begin{equation}
\sigma_f^2 =  g^T V g
\end{equation}

where $ \sigma_f^2 $ is the variance in the function $ f $ (i.e., Eq.~\eqref{eq:zpe} in the present case) of the set of parameters
$ Y_{\,l0} $, whose variance-covariance matrix is $ V $, with the $ i^{th} $ element in the vector $ g $ being
$ \dfrac{\partial f}{\partial Y_i} $ .

Discrepancies of $\sim 2$\,\wn are observed by comparing our data with those reported in 
Ref.~\citenum{Roueff2015} because their definition of the ZPE does not include the term 
$Y_{00}$, which is non-negligible for light molecules\cite{Irikura2007}.
These newly determined values should be used in the calculation of the exoergicity values 
$\Delta E$ of chemical reactions relevant in fractionation processes.

\begin{table}[!b]
 \caption[]{Zero Point Energies (in \wn) of imidogen isotopologues.
            \label{tab:zpe}}
 \centering\small
 \vspace{2ex}
 \begin{tabular}{r c r r}
  \hline \\[-1ex]
  Species    &  This work  &  Ref.~\citenum{Irikura2007} & Ref.~\citenum{Roueff2015} \\[0.5ex]
  \hline \\[-1.5ex]
  NH    & 1623.5359(17) & 1623.6(6)  & 1621.5 $^a$ \\
  \qnh  & 1619.9485(17) &            & 1617.9 $^b$ \\
  ND    & 1190.0859(11) & 1190.13(5) & 1189.5 $^c$ \\
  \qnd  & 1185.1413(11) &            & 1183.6 $^b$ \\
  \hline \\[-0.5ex]
 \end{tabular}
 \begin{minipage}{0.45\textwidth}
  \textbf{Notes.} \\
  Number in parentheses are the $1\sigma$ statistical errors in unit of the last quoted digit. 
  $^{(a)}$ From Ref.~\cite{Ram2010}. $^{(b)}$ Computed. $^{(c)}$ From Ref.~\cite{Dore2011}.
 \end{minipage}
\end{table}

\section{Conclusions}
\indent\indent
In this work the pure rotational spectrum of \qnd in its ground electronic $X^3\Sigma^-$ 
state has been recorded for the first time using a frequency-modulation submillimeter-wave 
spectrometer. 
A global fit, including all previously reported rotational and ro-vibrational data for the 
other isotopologues of the imidogen radical, has been performed and yielded a comprehensive
set of Dunham coefficients.
Moreover, the Born--Oppenheimer Breakdown constants have been determined for 13~parameters 
and also the adiabatic contribution of the terms $\Delta^N_{01}$ and $\Delta^H_{01}$ were 
evaluated and compared to theoretical estimates.
The present analysis enables to predict rotational and ro-vibrational spectra of any isotopic 
variant of NH at a high level of accuracy and to assist further astronomical searches of imidogen.
From our results, very accurate values of the equilibrium bond distances $r_e$ and the 
vibrational Zero-Point Energies for the different isotopologues have been derived.

\section*{Acknowledgements}
This work was supported by Italian MIUR (PRIN 2015 "STARS in the CAOS") and by the University of Bologna (RFO funds).


\bibliography{bis}
\bibliographystyle{rsc} 



\end{document}